\documentclass{aa} 
\usepackage{graphics}
\usepackage{psfig}
\def\la{\raise.5ex\hbox{$<$}\kern-.8em\lower 1mm\hbox{$\sim$}}
\def\ma{\raise.5ex\hbox{$>$}\kern-.8em\lower 1mm\hbox{$\sim$}}

\def\msol{M$_{\odot}$ }
\def\kms{$\rm km\, s^{-1}$}
\def\cm3{$\rm cm^{-3}$}
\def\Ts{$T_{\rm *}$}
\def\Vs{$V_{\rm s}$}
\def\n0{$n_{\rm 0}$}
\def\B0{$B_{\rm 0}$}

\def\erg{$\rm erg \, cm^{-2} \, s^{-1}$}

\def\Hb{H$\beta$}

\def\kms{$\rm km \, s^{-1}$}
\def\cm3{$\rm cm^{-3}$}
\def\Ts{$\rm T_{*}$}
\def\Vs{$\rm V_{s}$}
\def\n0{$\rm n_{0}$}
\def\B0{$\rm B_{0}$}

\def\Hb{H$\beta$}

\def\erg{$\rm erg \, cm^{-2} \, s^{-1}$}

\begin{document}

   \thesaurus{} 

\title{Internal and external shock fronts in  He2-104}

\author{
M. Contini\inst{1,2}
\and L. Formiggini\inst{3,4}
}
   
\offprints{M.\ Contini, contini@ccsg.tau.ac.il}

   \institute{
School of Physics \& Astronomy, Tel Aviv University, 
69978 Tel Aviv, Israel
\and 
Astrophysikalisches Institut Potsdam, An der Sternwarte, 16, D-14482
Potsdam, Germany
\and 
Wise Observatory, Tel Aviv University, 69978 Tel Aviv, Israel
\and
Istituto di Radioastronomia, CNR,via P. Gobetti 101, 40129 Bologna, Italy
}
   \date{Received ??; accepted ??}

\authorrunning{M.\ Contini and L.\ Formiggini}
\titlerunning{Shock fronts in He2-104}
   \maketitle

\begin{abstract}

The spectra  of both the inner and outer nebulae  in He2-104 are modelled
by composite models which accounts consistently for photoionisation and shocks.
The temperature of the hot star results  \Ts $\sim$ 130 000 K.
We suggest that  the characteristic "crab legs" in He2-104 are produced  
by R-T instability at the shock front.
The calculated X-ray flux is below the detection limit of ROSAT.

\keywords{shock waves;stars: binaries: symbiotic-stars:individual: He2-104}

\end{abstract}

\section{Introduction}

Symbiotic novae (SBN) are binary  systems which consist of a  white dwarf (WD) and a late type giant,
often a Mira variable.
Colliding winds from both components after  an outburst create  complex hydrodynamic 
conditions within the system and outside.
The importance of shock hydrodynamics  to the evolution of symbiotic
systems has been  demonstrated  by the morphological
development at later stages (Hack \& Paresce 1993, Contini 1997, Contini \& Formiggini 1999, etc.).
Particularly, new observational techniques  detected morphological structures extending up to 2 pc,
with complex phenomenology
(jets, expanding shells, "crab legs", etc.) in the circumstellar
material surrounding SBN. Actually, Corradi et al (1999a) detected an extended
optical nebula in 8 D-type (dust rich) SBN, about 40\%  of the observed sample.
The circumbinary gas was observed in the light of [NII] 6584, which is the
strongest line in most extended nebulae (Corradi et al 1999b).

In this paper we focus on He2-104  which  belongs to the 8 D-type SBN
in the Corradi et al sample.
The image  which  appeared in an HST News Release on August 1999
($\it http://oposite.stsci.edu/pubinfo/pr/1999/32/$) is presented in Fig. 1
(see also Corradi, 2000, Fig. 1).

\begin{figure}
\psfig{figure=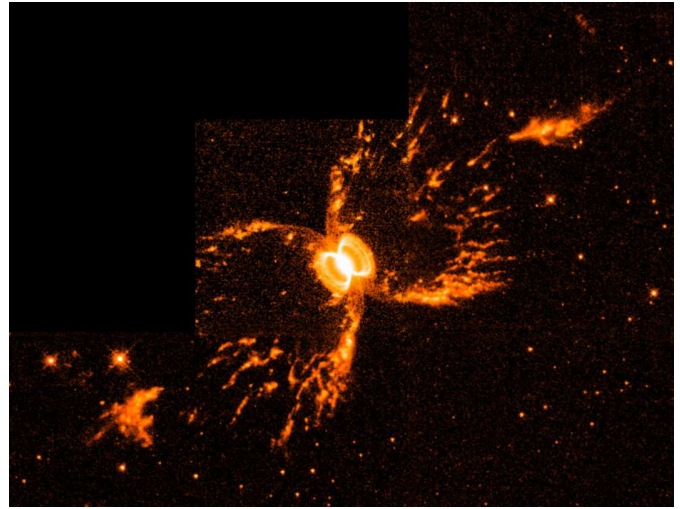,width=.49\textwidth}
  \caption{HST [NII] image of He 2-104
}
\label {fig:fus}
\end{figure}

The morphology of He2-104 is described by Corradi \& Schwarz (1993).
The line spectra emitted by the inner nebula of He2-104 have  
been  observed by  Lutz et al. (1989) and  Freitas Pacheco \& Costa (1996).
The line ratios indicate a large range of temperatures and densities.
Actually, Freitas Pacheco \& Costa "did not succeeded to reproduce the observed line ratios"
by pure photoionisation models and  suggested  a contribution from shocks.
Shocks were already invoked by Schwarz, Aspin, \& Lutz (1989) for He2-104.
In the present work we model the spectra by  composite models  which consistently account
 for the shock  as well as for the radiation from the hot star.
 The SUMA code (Viegas \& Contini 1994) is adopted.

Narrow band images in [OIII] 5007, H$\alpha$, [NII] 6584, and [SII] 6731
were  presented by Schwarz et al (1989), Lutz et al (1989), and Corradi \& Schwarz
(1993).
Detailed spectra of the outer nebula  are not yet available,
the only information being that [NII]  is the brightest line
(Corradi 2000). Nevertheless,
 the characteristics of the  nebula are investigated on the basis of the
qualitative information  derived from the narrow
band images.
We note that first ionisation level lines (e.g [NII]) 
are generally strong in shock dominated spectra.

Composite models are presented in Sect. 2.   He2-104 inner nebula spectra are modelled
in Sect. 3 and the outer nebula is discussed in Sects. 4 and 5.
Concluding remarks follow in Sect. 6.

\section{Composite models for the calculation of the spectra}

We have run a grid of composite models with the code SUMA (see Contini 1997). 
The input parameters are those referring to the hot star (the effective
temperature, \Ts, the ionisation parameter, U) and those which refer to the shock
(the shock velocity, \Vs, the preshock density, \n0, and the preshock
magnetic field, \B0). Moreover, the relative abundances to H of
He, C, N, O, Ne, Mg, Si, S, A, and Fe, and the
geometrical thickness, D, of the emitting filaments are also included.

Both Kaler (1976) and Kenyon \& Webbink (1984) suggest
that for very hot stars the blackbody approximation provides a better fit
to the nebular emission lines than the stellar models.
A black body approximation is adopted in SUMA.
Particularly, a grid of models 
with input parameters in the ranges suitable for SBN is presented
by Formiggini, Contini, \&  Leibowitz (1995, Table 3).
 
\section{The physical conditions in the inner nebula}

We  apply to He2-104 the same model  that was used  for modelling SBN previously
(Formiggini et al. 1995, Contini 1997, Contini \& Formiggini 1999),
i.e.  we assume that the  spectrum is emitted  from the interaction zone
of two colliding stellar winds, one from the WD and the other from the late giant.
Two shock fronts form at collision (Contini \& Formiggini 2000), one propagating in reverse 
towards the hot star and the other expanding outwards the system.
The former is responsible for emission of lines from relatively high ionisation levels,
the latter for low level and neutral lines. 

 The emitting nebulae are, therefore, heated and 
photoionised  both by the shock created by collisions and by the  radiation flux
from the hot star. In the reverse shock both the radiation and the shock act on the internal edge
of the emitting nebula. In the expanding shock they act
on opposite edges.

We refer to the  optical spectrum  observed and reddening corrected  
by Freitas Pacheco \& Costa (1996). 
The calibration of the optical spectra observed by Lutz et al. (1989) is uncertain and
the line fluxes are not reddening corrected, so they  will be used only to  further
 constrain the model.
In Table 1 the strongest lines (relative to \Hb) and
the best fitting models are presented. The parameters are chosen phenomenologically by the  fit
of calculated to observed line ratios. Three reverse models (models 1, 2, and 3) and one
 expanding model (model 4*) are selected and  the input parameters are listed 
at the bottom of Table 1. \B0 is 10$^{-3}$ gauss  for all the models.

\begin{table*}
\caption[Table 1]{The comparison of calculated with observed line ratios to \Hb=1}
\begin{flushleft}
\small{
\begin{tabular}{l l llll ll llll}     \\ \hline
\  line & obs$^1$ & model 1 & model 2 & model 3 & model 4*  & model SUM\\
\hline\\
\ [OII]3727 & 0.057  & 0.036 &  0.01  & 0.021 & 1.2 & 0.04 \\
\ [NeIII]3869+ & 1.9 & 2.27 &  2.77  & 2.38& 2.5 & 2.72 \\
\  [OIII]4363 & 0.72  & 0.38  & 0.83  & 0.52& 0.16 & 0.8\\
\  HeII 4686 & 0.185  & 0.170  & 0.188 & 0.165& 0.37& 0.189 \\
\  [OIII]5007+ & 4.0  & 4.16  &  4.16  & 4.24& 17.2& 4.38 \\  
\  [FeVI]5176& 0.08  & 0.01  & 0.024 & 0.12 &  -&- \\
\  HeI 5876 & 0.14 & 0.13 & 0.13 & 0.124 & 0.13& 0.13 \\
\  [FeVII]6086 & 0.004 & 0.001 & 0.005 & 0.003& - &-\\
\  [NII]6584+ & 0.3  & 0.4   & 0.23  & 0.27& 3.8 &0.3\\
\  [SII]6716 & 0.0084 & 0.009 & 0.0084 & 0.008 & -&- \\
\  [SII]6730 & 0.011 & 0.02 & 0.019 & 0.018 &- &- \\
\hline   \\
\  \Vs ~(\kms)  & - & 150 & 300 & 300 & 50 &-\\  
\  \n0 (10$^4$ \cm3)  &-& 7  & 7  &  5& 2&- \\
\  \Ts (10$^5$K) &-& 1.5 & 1.3  & 1.4 &1.4&- \\
\  U &-&  0.05  & 0.4 & 0.15& 0.005&- \\
\   D (10$^{15}$cm) &-& 0.67 & 4.47 & 8.&2&- \\
\hline
\end{tabular}}

$^1$ Freitas Pacheco \& Costa (1996)
\end{flushleft}
\end{table*}

We refer  to the observed   CIV 1550/HeII 1640 line ratio by Lutz et al. 
to constrain the model.
Both lines are strong and the frequencies close enough to depend
less dramatically on the reddening correction, even in the UV.
The  observed ratio,   $\sim$ 2  is fitted only by model 2  which  yields 1.6 for it,
whereas models 1 and 3
under-predict the ratios by a  factor of 4. Therefore, we adopt model 2, characterized
by U=0.4, as the best fitting model.
Notice that the reverse shock velocity of 300 \kms 
does not represent the velocity of the expanding matter
observed by Schwarz et al (1989), namely, -36 $\pm$ 18 \kms in the northern lobe,
-139 $\pm$ 12 \kms in the central bright region, and -235 $\pm$ 15 \kms in the southern lobe. 

The calculated [OII]/\Hb ~is, however, much smaller than the observed one in model 2, as
strong lines from low ionisation levels are generally emitted by the expanding
shock (see e.g. Contini 1997 and Contini \& Formiggini 1999). This is represented by model 4*.
The presence of two shocks is generally revealed by the line profiles (Contini 1997).
Since we have no information from the data, we have  adopted
the expanding model  presented by Contini \& Formiggini (1999, Table 3, model exp3) for 
the SBN RR Tel.
The \Hb ~absolute flux  is 386 \erg and 0.84 \erg for models 2 and 4*, respectively.
The weighted sum of the  spectra shows that the best fit is obtained by the weight ratio 1 : 10
for model 2 : model 4*. The results are given in the last column of Table 1. 

R$_{OIII}$$\equiv$ [OIII] 5007/[OIII] 4363 is in good agreement with the observed value.
R$_{OIII}$ depends strongly on the electron density and temperature of the emitting gas.
High electron densities  follow from compression downstream and high temperatures
follow from  the relatively high shock velocity. 

The [HeII]/\Hb ~line ratio, on the other hand, is determined  particularly by \Ts. 
 The results of Table 1 indicate that the color
temperature of the hot star is high enough (\Ts=130 000 K) to unambiguously classify 
He2-104 as a symbiotic nova.

The  relative abundances   adopted  in these calculations are
 He/H=0.1, N/H=1.5$\times10^{-4}$, and 
O/H=4.6$\times$10$^{-4}$ (Formiggini et al. 1995).

The [FeVI] 5176/\Hb ~line ratio  is under-predicted by model 2
but overpredicted by model 3,
 while [FeVII]/\Hb ~is rather well fitted. Therefore, a model 
 between  models 2 and 3
would better fit the [FeVI]/\Hb ~line ratio. 
In this case the fit of all the line ratios would improve, except for
CIV 1550/HeII 1640.

The [SII] 6716/6730 line ratio is  under-predicted by model 2, indicating higher 
electron densities in the emitting gas. Freitas Pacheco \& 
Costa find a density stratification from different line ratios,
the density value deduced from the [SII] 6716/6730 ratio being the lower limit. 
A stratification of densities and
(temperatures) downstream is characteristic of models which account for the shock.
Moreover, the error of the observed lines is not quoted by Freitas Pacheco \& Costa.
A line intensity error of $\sim$ 20 \%  would improve the fit of the predicted ratio.

\section{The crab-like outer nebula}

In spite of the lack of quantitative data about the spectrum,
some general considerations can be derived.

The emitting filaments (or the  "crab legs", see Sect. 5) 
propagate outwards and   a shock front
develops on the outer edge. Radiation from the hot star reaches the inner edge
of the filaments.
The temperature of the hot star  (\Ts $\sim$ 130 000 K)  results  from detailed 
modelling of the inner nebula (Sect. 3).
An effective temperature of  \Ts=130 000 K, implies
an ionising photon flux of $\cal N$=3.3 10$^{26}$ photons $\rm cm^{-2} s^{-1}$.
This flux is related to the ionisation parameter, U, and to the gas (number) density 
downstream, n, in the radiation dominated region of the filaments, 
by : $\cal N$ $\rm (R_{WD}/r)^2$ = U n c, 
where $\rm R_{WD} \sim 10^9$ cm is the radius of the WD, and r $\sim$ 0.3 pc 
(Corradi 2000) is the distance of the filament region from the hot source. We obtain :

\bigskip

\noindent
 U = 0.01(\Ts/1.3 10$^5$K)$^3$($\rm R_{WD}$/10$^9$cm)$^2$(r/10$^{18}$cm)$^{-2}$/n 

\bigskip

The downstream compression (n/\n0) increases with higher \Vs ~and decreases with higher
\B0. Even  adopting for \n0 the ISM density value  (\n0$\geq$1 \cm3) 
the resulting n would lead to a very low U. 
This is reasonable in view of the large distance of
the  filaments from the hot source. 
The gas inside the filaments  will be mainly ionised and heated by the shock.
Consequently, shock dominated (U=0) models  will be adopted
in the calculation of line intensities.

The  models selected  from the grid, which  show the observed spectrum characteristics
(prominent [NII] 6584 line absolute and 
relative to [OIII] 5007) are presented in Table 2.  The
input parameters appear in columns 2-5, and in columns 6-9  the absolute fluxes of the [OIII] 5007,
[OII] 3727, [NII] 6584, and \Hb ~lines are presented.
The observed expansion velocities of
the outer nebula are about 50-100 \kms (Corradi \& Schwarz 1993) and  $\sim$ 250 \kms
in the blobs related  to the jets (Corradi \& Schwarz 1995).
So, we present in Table 2 the results of calculations for \Vs ~between 50 and 250 \kms.

\begin{table*}
\caption[Table 2]{The line fluxes calculated at the emitting nebula (in \erg)}
\begin{flushleft}
\small{
\begin{tabular}{ l l l l l l l l l l l } \\ \hline\\
       &  \Vs   & \n0   & \B0 &   D   &  F$_{[OIII]}$ &F$_{[OII]}$ & F$_{[NII]}$  & F$_{H\beta}$\\
      &   (\kms) & (10$^4$ \cm3)  & (10$^{-5}$ gauss) & (10$^{16}$ cm) & && &  \\ 
\hline\\
 model O1   & 50    & 1   & 1   & $>$1& 0.00024 & 0.029  & 0.048  & 0.0046 \\
 model O2   & 50   & 1    & 10   & 0.1& 0.00024 &0.029    & 0.047 & 0.0044 \\
 model O3   & 50   & 2    & 1   & 1& 0.0004 & 0.03    & 0.07  & 0.0093  \\  
 model O4  & 100   & 1    & 1   &2.5&0.15  &  0.049   & 0.24  & 0.49   \\
 model O5  & 150   & 0.1    & 1   & 1 & 0.0027& 0.012  & 0.014 & 0.00167 \\
 model O6 & 250   & 0.1&  100   & $>$1 & 0.08  & 0.33 &   1.1   & 0.096 \\
\hline\\
\end{tabular}}
\end{flushleft}
\end{table*}

In Fig. 2 the physical conditions of the emitting gas relative to models O1 and O6 (Table 2),
which show  extreme \Vs, are presented. 

\begin{figure}
\psfig{figure=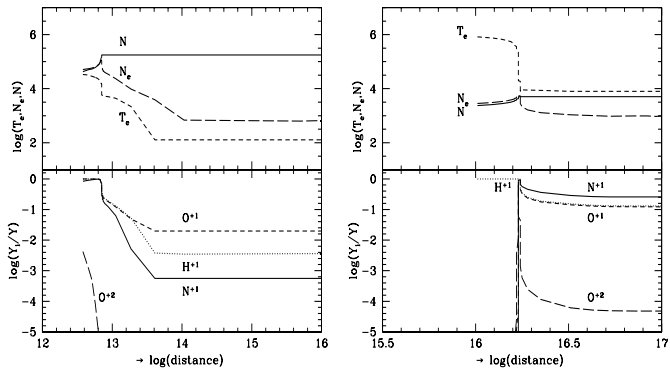,bbllx=125pt,bblly=280pt,bburx=800pt,bbury=490pt,width=17.5cm,clip=}
  \caption{
Model O1 left, model O6 right.  Top panels: The profiles of the electron temperature (short dashes),
the electron density (long dashes), and of the  density (solid line)
 as function of distance
from the shockfront (on the left side of each diagram).
Bottom panels:
The profiles of the fractional abundances of the most significant ions
(N$^{+1}$: solid line,
O$^{+1}$: short-dashed line,
H$^{+1}$: dotted line,
O$^{+2}$: long-dashed line)
as function of distance from the shockfront.}
 \label{fig:hock}
\end{figure}

The electron densities downstream in both  diagrams (Fig. 2) are in agreement with the
densities of 300-1000 \cm3 found in the "crab legs" by
Corradi \& Schwarz (1993).
Interestingly, for model  O1 (Fig. 2 left)
N$^{+1}$  is  lower than O$^{+1}$ , while F$_{[NII]}$ is higher than F$_{[OII]}$.
 In fact,
both the ionisation potential, $\chi$,  and the collision strength, $\Omega$,  
are important  in the calculation of collisional excited lines : 
F$_{line}$ $\propto$ $\Omega$ exp(-$\chi$) Y$_i$/Y (Osterbrock (1989).
Actually, $\chi$ is 1.90 eV and 3.32 eV for the 
transitions  $^3$P - $^1$D (NII) and  $^4$S - $^2$D (OII),
respectively, and $\Omega$  is 2.99 and 1.47, respectively, 
leading to the intensities shown in Table 2.

The  relative high \Vs ~(250 \kms) in model O6 creates a large zone of high temperature 
gas ($\leq 10^6$ K), 
 which  corresponds to soft X-ray emission  from the postshock region of the blobs.
 However, this model refers to the jets, which angle is very small (see Fig. 1).
 The calculated X-ray flux  is 0.036 \erg at the nebula
  and $\ll$10$^{-13}$ \erg at earth, if the jet projected opening angle is $<$ 5$^o$.
Since the angle is $\ll$ 5$^o$, He2-104 flux is below
 the ROSAT all-sky- survey detection limit ($\sim$ 10$^{-13}$ \erg).

\section{The "crab legs"}

The outer nebula filaments, named "crab-legs", are clearly identified in Fig. 1.
The R-T instability created by turbulence at the shock front
leads to fragmentation of matter. 
The  strong resemblence between the
"crab legs" which appear in the He2-104 image (Fig. 1) and the boundary of
the shock front presented by Graham \& Zhang (2000)  and, particularly, by Gull (1975, Fig. 3 ii)
suggests that the "legs" are actually formed by R-T instability.

A rough evaluation of the filling factor can be derived from some general considerations.
Adopting for He2-104 the typical Mira mass loss rate, $\dot{M}$ = 10$^{-6}$ \msol yr$^{-1}$, 
and an age  t = 900 yr (Corradi 2000),
the ejected mass in He2-104 is M $\sim$  10$^{-3}$ \msol, 
The radius of the He2-104 outer nebula is about 0.3 pc (Corradi 2000).
The swept up mass to
ejected mass ratio is about 7, if the  number density  in the ISM is  2 \cm3.
Assuming that all the ejected+swept up matter resides in the "crab legs",
the filling factor, f, in the  outer nebula can be calculated by
 M(1+7)=4 $\pi$ r$^2$ dd m$_H$ n$_{leg}$ f, where dd is the 
geometrical thickness of the  region containing most of the "legs". 
We roughly derive from the image in Fig. 1 that dd 
is of the order of $\sim$ 3.5 10$^{17}$ cm.
Adopting  M=2 $\times 10^{30}$g, r= 10$^{18}$cm, 
and a density in the "legs" n$_{leg}$=1000 \cm3, the filling factor f results $\sim$ 0.002,
in agreement with the clumpy structure of the outer nebula. 

\section{Concluding remarks}

In this paper we have modelled the  spectra emitted  from both the inner   
and the outer  nebulae around He2-104.
The results of model calculations show the important role of the shocks.
 It is found that in the inner nebula   the preshock density 
is \n0= 7 10$^4$ \cm3. In the  outer nebula \n0= 10$^3$-10$^4$ \cm3.
The high gap between the density in the outer nebula "crab legs" and that in the
ISM,  shows that the characteristics of the original ejected matter,
even if partly washed out by merging with the ISM,
 are still recognizable. The ionisation parameter is U=0.4 in the inner nebula
and negligible in the  outer one, indicating that the effect of shocks dominates
far away from the  symbiotic system.
Shock velocities of 50 - 250 \kms  provide the  prevailing [NII] line in the outer nebula.
Soft X-ray emission is predicted from the jet blobs which show the highest shock velocity.
The magnetic field is \B0=10$^{-3}$ gauss in the inner nebula and 10$^{-3}$ - 10$^{-5}$ 
gauss in the outer nebula.
Finally, the temperature of the hot star which results from modelling is \Ts $\sim$ 130 000 K.

We suggest that the formation of the characteristic "crab legs" is  due to R-T
instability at the shock front.

More accurate and detailed results will be obtained by  modelling when  quantitative spectra 
rich in number of lines will be available from  observations.

\begin{acknowledgements}
We are grateful to D. Prialnik for  precious comments, and to the referee, H. E. Schwarz,
for helpful criticism.

\end{acknowledgements}

\end{document}